\documentclass[sigconf]{acmart}

\AtBeginDocument{%
  \providecommand\BibTeX{{%
    \normalfont B\kern-0.5em{\scshape i\kern-0.25em b}\kern-0.8em\TeX}}}

\usepackage{amsmath,amssymb,amsfonts}
\usepackage{graphicx}
\usepackage{textcomp}
\usepackage{xcolor}
\usepackage{hyperref}       
\usepackage{url}            

\usepackage{booktabs}       
\usepackage{amsfonts}       
\usepackage{nicefrac}       
\usepackage{microtype}      
\usepackage{multirow}
\usepackage{subcaption}
\usepackage{tabularx}
\usepackage[linesnumbered,noend]{algorithm2e}
\usepackage{enumitem}
\usepackage{cleveref}
\usepackage{array}

\usepackage{dcolumn}
\usepackage[title]{appendix}

\newcolumntype{d}{D{.}{.}{2}}

\newcommand{\topic}[1]{}

\newcommand{\toolname}{\texttt{RecRec}\xspace}

\DeclareMathOperator*{\argminA}{arg\,min}
\DeclareMathOperator*{\argmaxA}{arg\,max}
\DeclareMathOperator*{\argsortA}{arg\,sort}

\newcommand{\norm}[1]{\left\lVert#1\right\rVert}
\usepackage[compact]{titlesec}   
\copyrightyear{2023}
\acmYear{2023}
\setcopyright{rightsretained}
\acmConference[CIKM '23]{Proceedings of the 32nd ACM International Conference on Information and Knowledge Management}{October 21--25, 2023}{Birmingham, United Kingdom}
\acmBooktitle{Proceedings of the 32nd ACM International Conference on Information and Knowledge Management (CIKM '23), October 21--25, 2023, Birmingham, United Kingdom}\acmDOI{10.1145/3583780.3615181}
\acmISBN{979-8-4007-0124-5/23/10}

\ccsdesc[500]{Information systems~Recommender systems}
\ccsdesc[300]{Computing methodologies~Machine learning}

\begin{document}

\title{\texttt{\toolname}: Algorithmic Recourse for Recommender Systems}


\author{Sahil Verma}
\email{vsahil@cs.washington.edu}
\affiliation{
\institution{University of Washington}
  \city{Seattle}
  \state{WA}
  \country{USA}
}


\author{Ashudeep Singh}
\email{ashudeepsingh@pinterest.com} 
\affiliation{
\institution{Pinterest, Inc.}
  \city{San Francisco}
  \state{CA}
  \country{USA}
}

\author{Varich Boonsanong}
\email{varicb@cs.washington.edu} 
\affiliation{
\institution{University of Washington}
  \city{Seattle}
  \state{WA}
  \country{USA}
}

\author{John P. Dickerson}
\email{john@cs.umd.edu}
\affiliation{
\institution{University of Maryland}
  \city{College Park}
  \state{MD}
  \country{USA}
}

\author{Chirag Shah}
\email{chirags@uw.edu}
\affiliation{
\institution{University of Washington}
  \city{Seattle}
  \state{WA}
  \country{USA}
}

\begin{abstract}
  Recommender systems play an essential role in the choices people make in domains such as entertainment, shopping, food, news, employment, and education. 
The machine learning models underlying these recommender systems are often enormously large and black-box in nature for users, content providers, and system developers alike. 
It is often crucial for all stakeholders to understand the model's rationale behind making certain predictions and recommendations. 
This is especially true for the content providers whose livelihoods depend on the recommender system. 
Drawing motivation from the practitioners' need, in this work, we propose a recourse framework for recommender systems, targeted towards the content providers. 
Algorithmic \emph{recourse} in the recommendation setting is a set of actions that, if executed, would modify the recommendations (or ranking) of an item in the desired manner. 
A recourse suggests actions of the form: ``if a feature changes $X$ to $Y$, then the ranking of that item for a set of users will change to $Z$.'' 
Furthermore, we demonstrate that \toolname is highly effective in generating valid, sparse, and actionable recourses through an empirical evaluation of recommender systems trained on three real-world datasets. 
To the best of our knowledge, this work is the first to conceptualize and empirically test a generalized framework for generating recourses for recommender systems. 

\end{abstract}

\keywords{{\it Algorithmic Recourse, Recommender Systems, Explainable Recommender Systems}}

\maketitle


\section{INTRODUCTION}
\label{sec:intro}

Recommender systems are ubiquitous in online platforms today. They have a major influence on our choices in domains ranging from entertainment, social media, and shopping to news and education. 
These systems operate by filtering items from a large set to provide the most relevant ones to the user. 
Recommender systems can broadly be classified into two categories, content filtering and collaborative filtering \citep{Adomavicius2005Toward,recobroad1,recobroad2,recobroad3}. 
Content filtering represents each item using a set of features and recommends items to users based on the similarity to items consumed by the user in the past \citep{content1,content2,content3}. 
For example, if a user has purchased a cell phone recently, a content filtering system would recommend a phone case to the user based on its similarity to the phone. 
Collaborative filtering-based recommender systems recommend items to a user based on other users' interests who have a similar user history. 
For example, if a user recently bought a cell phone, a collaborative filtering-based recommender system would recommend the user with a phone case based on the information that other users who bought a cell phone also bought a phone case in the past.

Modern recommender systems are black-box models that has spurred the development of inherently interpretable recommender models or techniques to explain the factors influencing the recommendations \citep{survey-reco-xai}. 
We have also seen the rapid adoption of incorporating explainability in real-world recommender systems. 
Facebook offers ``Why am I seeing this ad?'' for every sponsored advertisement on its platform, 
and Amazon offers reasons why a product is recommended to you tab. These explanations are broadly termed as {\em feature attribution explanations} as they highlight a part of the features that lead to the recommendations. 


All these explanations are primarily geared towards the end-users of the recommender platforms 
However, recommender systems are usually multi-stakeholder platforms with content providers and the system developers' interests baked into the system \citep{multistakeholder-reco1,multistakeholder-reco2,multistakeholder-reco3}. And these stakeholders also need transparency into the system. 
Since the content providers are dependent on the platform for their livelihood, they are interested in understanding the factors that 
influence their product's rank in the recommendations \citep{airbnbhosts-course,Upwork-course1,Upwork-course2}. 
There have been several studies to understand the perspective of content providers offering services on several kinds of such platforms. 
\citet{airbnbhosts-xaineed} did a study with several Airbnb hosts to understand their perspectives. \citet{Upwork-xaineed} interviewed freelancers working on Upwork. \citet{Etsy-xaineed} interviewd sellers on handmade product platform Etsy.  
Most content providers expressed the helplessness they face in understanding the factors that influence 
their product's rank on the platform \citep{airbnbhosts-xaineed,Upwork-xaineed,Etsy-xaineed,FB-seller-xaineed}, and would like to gain transparency into it (see \Cref{sec:need-for-recourse}). 

The kind of explainability that the content providers seek is similar to counterfactual explanations in classification systems \citep{wachter2018counterfactual,verma2020counterfactual}. 
In general, counterfactual explanations describe a causal situation of the form: `If I change X to Y, the outcome will change to Z'. 
Counterfactual explanations are frequently employed to answer questions like ``What change in my features would help me to get the loan?''. 
In recommender systems, a set of factors that improves the rank of a product offer a causal explanation of what changes would lead to an improved ranking for it. 
This set of factors are termed as {\em recourse} if the content providers can alter them to improve their product's rank. 
Since studies have shown that providing transparency into the algorithmic system improves the user's trust and adoption of the platforms \citep{reco-xai-trust}, the motivating reason for incorporating counterfactual explanations in recommender systems is even more compelling. 


The primary contribution of our work is to conceptualize a generalized framework for generating algorithmic recourse-based explanations for recommender systems. 
In this work, we use the terms counterfactual explanations and algorithmic recourse interchangeably, as we are using counterfactuals to provide recourses to the content providers, system developers, or curious end-users of a recommender system. 
Under this framework, we propose a novel algorithm, \toolname, that generates algorithmic recourses for a real-world recommender systems. 
\toolname casts the problem of finding recourses as an optimization problem that is solved using gradient descent in the feature space. 

We first establish the desirable properties of a recourse in recommender systems (\Cref{sec:desiderata}) and then cast the recourse generation problem as an optimization problem (\Cref{sec:algorithm}), which \toolname solves. 
We conduct experiments with three different recommender systems 
and show \toolname's efficacy in generating recourses that achieve high success rate while satisfying the other desirable properties (\Cref{sec:evaluation}). 
We make the following key contributions:
\begin{enumerate}[leftmargin=9pt,topsep=0pt,partopsep=0pt]
    \item We establish the desirable properties for algorithmic recourses in recommender systems (\Cref{sec:desiderata}). 
    \item We propose a novel approach called \toolname to generate recourses for a broad class of recommender system architectures (\Cref{sec:algorithm}). 
    \item We empirically demonstrate the effectiveness of \toolname through extensive experiments on three recommender systems trained on real-world datasets (\Cref{sec:evaluation}). 
\end{enumerate}

\section{DESIRABLE PROPERTIES OF ALGORITHMIC RECOURSE}
\label{sec:desiderata}

To be effective for a content provider, a recourse in a recommender system setting should satisfy several desirable properties:
\begin{enumerate}[leftmargin=4pt,topsep=0pt,partopsep=0pt]

\item {\it Valid: } A recourse when executed should lead to an increased exposure for the concerned item among the users of the target group, and therefore should have an improved rank after the recourse. 
 
 \item{\it Sparse Changes: } A recourse should not change many features of the item. Being close to the original features makes the recourse more easily achievable \citep{Miller-xai:2019}. 
 
 \item {\it Minimal side-effect }: A recourse should ideally only move the concerned item to an improved rank and have minimal side-effect on the ranks of the other items \citep{eval-recsys} (specially near the top ranks). 

\end{enumerate}

\section{\toolname's ALGORITHM TO GENERATE RECOURSE}
\label{sec:algorithm}

This section formulates the recourse from the perspective of a content provider wanting to change the features of an item so that it gets more exposure for users in a target group, i.e., within the top-$k$ recommendations for these users. 

Given an item's original features, $r$, the goal is to find updated features $r'$, such that the item, \emph{item}, is recommended within the \emph{top-$k$} ranks for a group of target users
In content filtering based systems, the features of an item are its attributes that are used to measure the similarity between different items, e.g. genres of a movie or book. 
\Cref{tab:notations} lists the notation used in this section.

\begin{table}[h]
    \centering
        \caption{\small Notations for \toolname's Algorithm}
        \label{tab:notations}
    \resizebox{0.85\columnwidth}{!}{
    \begin{tabular}{p{0.6in}p{2.6in}}
    \toprule
    \textbf{Notation} & \textbf{Description} \\ \midrule
        U: & Set of users \\
        I: & Set of items \\
        $\mathcal{I}$: & $\{v_i: i \in I, v_i \in \mathbb{R}^f \}$, $v_i$ is the item features for item i. \\
        $\mathcal{R}_j \in \mathbb{R}^{|I_j|}$:  & $\mathcal{R}_j[k]$ denotes the rating given by user j to item $I_j[k]$ \\
        S: & Set of target users \\
        a: & The item for which recourse is being sought \\ 
    \bottomrule
    \end{tabular}
    }
\end{table}

\SetKwProg{Fn}{Function}{}{}
\RestyleAlgo{algoruled}
\newcommand\mycommfont[1]{\tiny\ttfamily\textcolor{blue}{#1}}
\SetCommentSty{mycommfont}

\newcommand{\xAlCapSty}[1]{\small\sffamily\bfseries\MakeUppercase{#1}}
\SetAlCapSty{xAlCapSty}


\newcommand\mynlfont[1]{\scriptsize\sffamily{#1}}
\SetNlSty{mynlfont}{}{}
\SetAlFnt{\scriptsize\sf}

\SetSideCommentLeft

\begin{algorithm}
 \DontPrintSemicolon 
 \SetKwInOut{KwIn}{Input}
 \SetKwInOut{KwOut}{Output}
 \SetKwData{desiredrank}{DesiredRank}
 \SetKwData{lr}{LearningRate}

 \KwIn { Item features $\mathcal{I}$, the target user group S, ratings given by each user $\mathcal{R}$, the concerned item a, the desired rank of the item (e.g. top-10), hyperparameter $\lambda$, hyperparameter \lr}
 \KwOut{ New item feature for concerned item a: $v_{a}^*$}

 \caption{ \label{alg:recourse_content} \small Generate recourse to move an item to the \emph{top-k} recommendations for a target group of users.} 
 
 \SetKwFunction{computerec}{Compute\_Recourse}
 \SetKwData{loss}{loss}
 \SetKwData{iters}{iterations}
 \SetKwData{maxiters}{maxiterations}
 \SetKwFunction{graddescent}{gradient\_descent}
 \SetKwFunction{getupdatedrank}{get\_updated\_ranks}
 \SetKwData{newrank}{newrank}
 \SetKwData{newscore}{newscore}

    \Fn{$\mathit{\computerec(\mathcal{I}, S, \mathcal{R}, a, \desiredrank, \lr, \lambda)}$}{
        $\mathit{\iters \leftarrow 0}$ \\
        $\mathit{v_a, v_a' \leftarrow \mathcal{I}[a]}$ \\
        $\mathit{S' \leftarrow S}$ \\
        \While{$\mathit{\iters < \maxiters}$ }{ \label{line:gdstart}
            $\mathit{\loss \leftarrow -\big( \sum_{j \in S'} v_a \mathcal{I}_j^\top \odot \mathcal{R}_j \big)  \ + \lambda * \norm{v_a' - v_a}_1 }$ \\
            \tcp{In each iteration, perform a gradient descent step for the \loss}
            $\mathit{v_a' \leftarrow \graddescent(\lr, \loss)}$ \label{line:newrat}  \\ 
            \tcp{Get the updated rank of the item for each user in S}
            $\mathit{\newrank \leftarrow \getupdatedrank(\mathcal{I}, v_a', \mathcal{R}, S)}$
            \tcp{Remove users from S for whom the concerned item is within the desired rank}
            \For{$\mathit{u \in S}$}{  \label{line:onlynotyetuser}
                \If{$\mathit{\newrank[u] \in \desiredrank}$}{
                    $\mathit{S' \leftarrow S \setminus u}$
                }
            }
            \label{line:gdend}
        }
        \Return{$\mathit{v_a'}$}
    }
    

    \Fn{$\mathit{\getupdatedrank(\mathcal{I}, v_a', \mathcal{R}, S)}$}{
        \tcp{Compute the new score of each item for each user}
        \For{$\mathit{u \in S}$}{
            $\mathit{\newscore \leftarrow \emptyset}$ \\
            \For{$\mathit{z \in \mathcal{I}}$}{
                $\mathit{\newscore[z] \leftarrow \sum v_z \mathcal{I}_j^\top \odot \mathcal{R}_j}$
            }
            \tcp{sort the $\newscore$ in descending order to generate new ranks}
            $\mathit{\newrank[u] \leftarrow \argsortA_{z \in \mathcal{I}} \newscore[z]}$
        }
        \Return{$\mathit{\newrank}$}
    }

\end{algorithm}

\Cref{eq:content_recourse} shows the objective function that we need to optimize to generate a recourse that achieves the desired change in the rank of the item while not changing its feature too much. 

\begin{equation}
    \begin{aligned}
        v_a^* = \argmaxA_{v_a'} & \sum_{j \in S} v_a' \mathcal{I}_j^\top \odot \mathcal{R}_j \hspace{2.5em}
    s.t. \norm{v_a' - v_a}_0 \leq \epsilon
    \end{aligned}
    \label{eq:content_recourse}
\end{equation}

We use gradient descent to optimize the objective function for which we need encode the twofold goal as a constrained optimization problem with sparsity inducing L1 norm as the constraint (with $\lambda$ as the hyperparameter). 

\begin{equation}
    \argminA_{v_a'} -\big( \sum_{j \in S} v_a \mathcal{I}_j^\top \odot \mathcal{R}_j \big)  \ + \lambda * \norm{v_a' - v_a}_1
    \label{eq:content_recourse_final}
\end{equation}

\Cref{alg:recourse_content} provides the algorithm to generate recourses for content filtering based recommender systems. 
The algorithm takes as input the features of all item $\mathcal{I}$, the target user group $S$, the ratings given by the users $\mathcal{R}$, the item for which a recourse is desired $a$, and the desired rank for the item $DesiredRank$. 
It runs the gradient descent algorithm until convergence. 
In each iteration, the algorithm computes the loss given in \cref{eq:content_recourse_final} and updates the features of the item using the gradient descent algorithm. 
The first term of the loss function is only computed for the subset of users in the target group $S$ for whom the concerned item has not yet been ranked within the desired rank (\cref{line:onlynotyetuser}). This helps in two ways:
a) encourages the algorithm to change the item features to move the item within the desired rank for a larger number of users from the target group, and
b) limits the change in the users' recommendation lists (which is another desirable property of a recourse). 

\paragraph{Iterative Hard Thresholding}
We use L1 norm to induce sparsity in the change between the original and the recourse item features. However, this might not be sufficient to ensure that the recourse is sparse. Therefore, after the convergence of \cref{alg:recourse_content}, we iteratively set the values of the features with the smallest absolute difference with the original features to the original feature value at those indices, a process termed as iterative hard thresholding \citep{iterative-hard-thr}. 
This leads to a tradeoff between the number of users for whom the item is moved within the desired rank (\textit{success rate}) and the sparsity of the recourse. We continue iteratively hard thresholding until we start losing more than a certain percentage of the success rate (in our experiments we set this threshold to 20\%).

\section{EVALUATION}
\label{sec:evaluation}

We performed experiments using three real-world recommender systems to measure \toolname's efficacy, efficiency, and side-effect when generating recourses for items at various ranks:


\begin{figure*}
    \centering
    \includegraphics[width=0.24\textwidth]{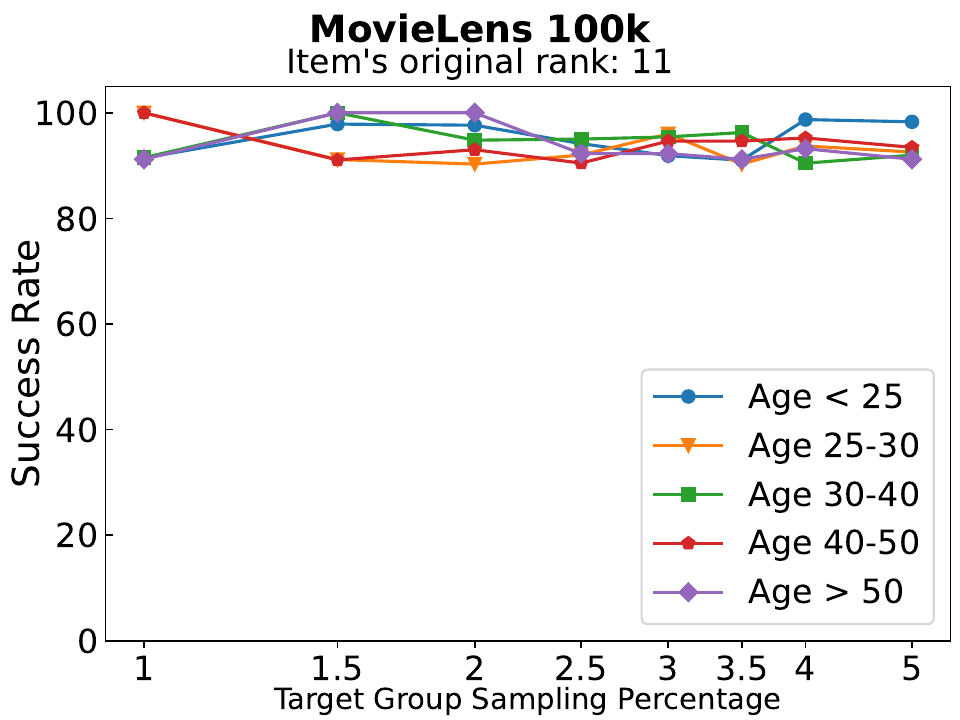}
    \hfill
    \includegraphics[width=0.24\textwidth]{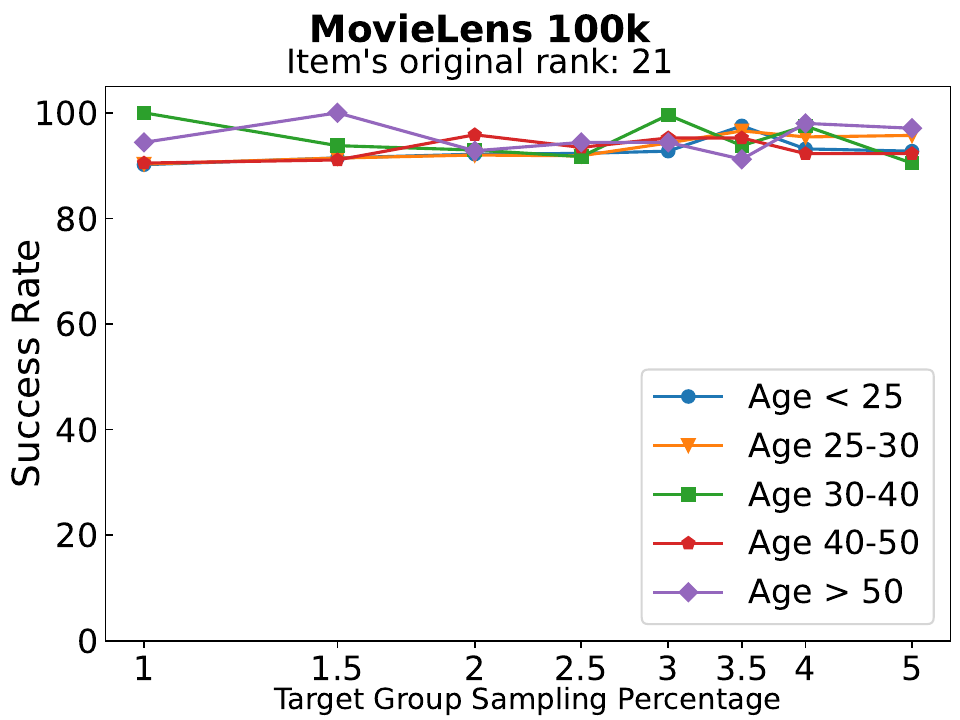}
    \hfill
    \includegraphics[width=0.24\textwidth]{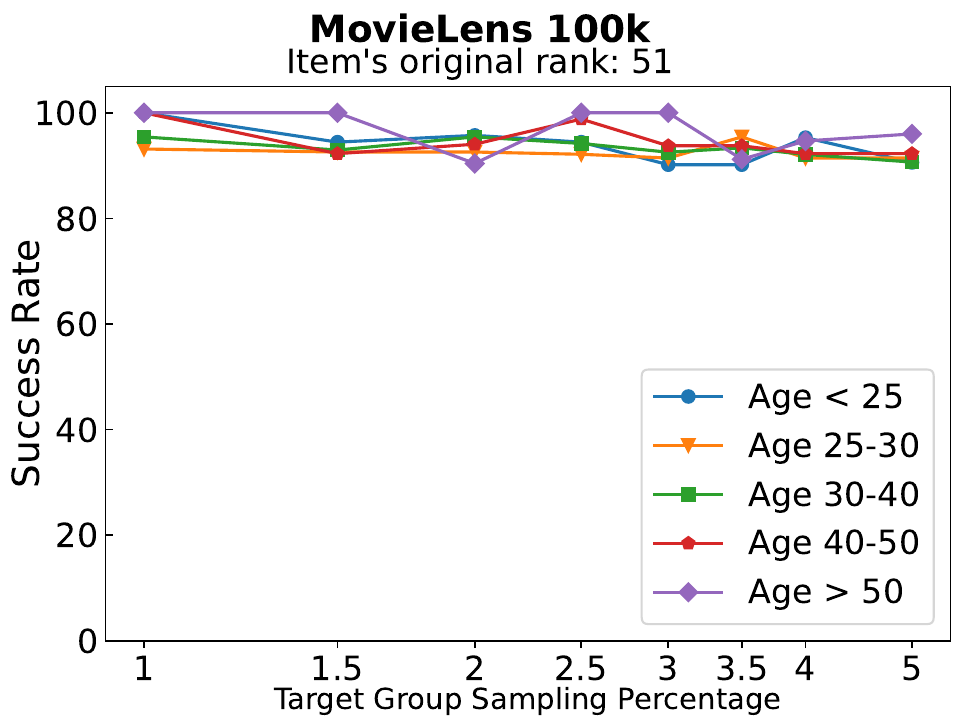}
    \hfill
    \includegraphics[width=0.24\textwidth]{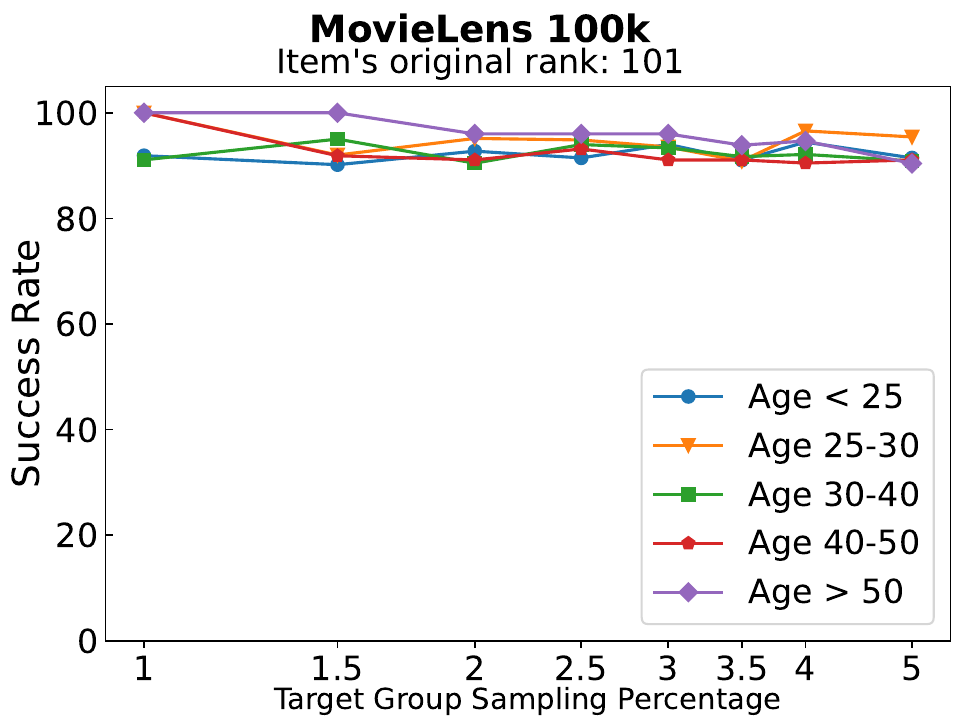}
    \caption{\small \toolname's success rate for items at original ranks 11, 21, 51, and 101 for the recommender system trained on MovieLens-100K. \toolname gets 100\% success rate for all user groups and items at all original ranks with very small sampling size starting from 1\% of the user group.}
    \label{fig:ml_100k_content_recsys}
\end{figure*}

\begin{enumerate}[leftmargin=0pt,topsep=0pt,partopsep=0pt]
    \item \textit{MovieLens-100K \citep{movielens-dataset}: } This movie recommendation dataset has about 1000 movies and 1700 users who gave a total of 100K ratings. Each movie has information such as its summary, actors, directors, and genre. We consider the movie summary as mutable and others as immutable features. Using standard NLP data processing we featurize each summary in about 9500 dimensions. Dot product between the feature vectors of two movies determine their similarity. If two movies are very similar and a user has liked one of them, the recommender system will recommend the other. MovieLens also provides the rating a user has given to certain movies which we use to weigh similarity between movies for providing more accurate recommendations. The dataset also provides user metadata, like age and occupation. In experiments, we use age ranges to group users into 5 equi-sized groups that content providers want to target. 
    
    \item \textit{AliEC Ads \citep{aliec-dataset}: } This dataset is an ad click prediction dataset provided by Alibaba. It has about 5600 ads and 13200 users who interacted with 1.4 million ads. For each ad, it provides information like its category, brand, and price. All features are considered mutable. We one-hot encode all categorical features and use price after normalization to feature each ad in about 2300 dimensions. Again, the dot product between feature vectors of two ads determine their similarity. 
    The dataset also provides user metadata such as age level and gender. In experiments, we use age level to group users into five equi-sized groups that content providers want to target. 
    
    \item \textit{Goodreads \citep{goodreads-dataset1,goodreads-dataset2}: } This book recommendation dataset has about 4300 books and 11200 users who have a total of 1.3 million ratings. Each book has features like its short description, genre, number of reviews, hardcover or ebook format. All features are considered mutable. Using standard NLP data processing, we featurize each movie description and other features in about 17400 dimensions. 
    The dot product measures the similarity between two books, and user ratings are used to weigh similarity. The dataset does not provide user metadata, and therefore we group users into five equi-sized groups based on the number of ratings they provided. 
\end{enumerate}

\begin{figure*}
    \centering
    \includegraphics[width=0.5\columnwidth]{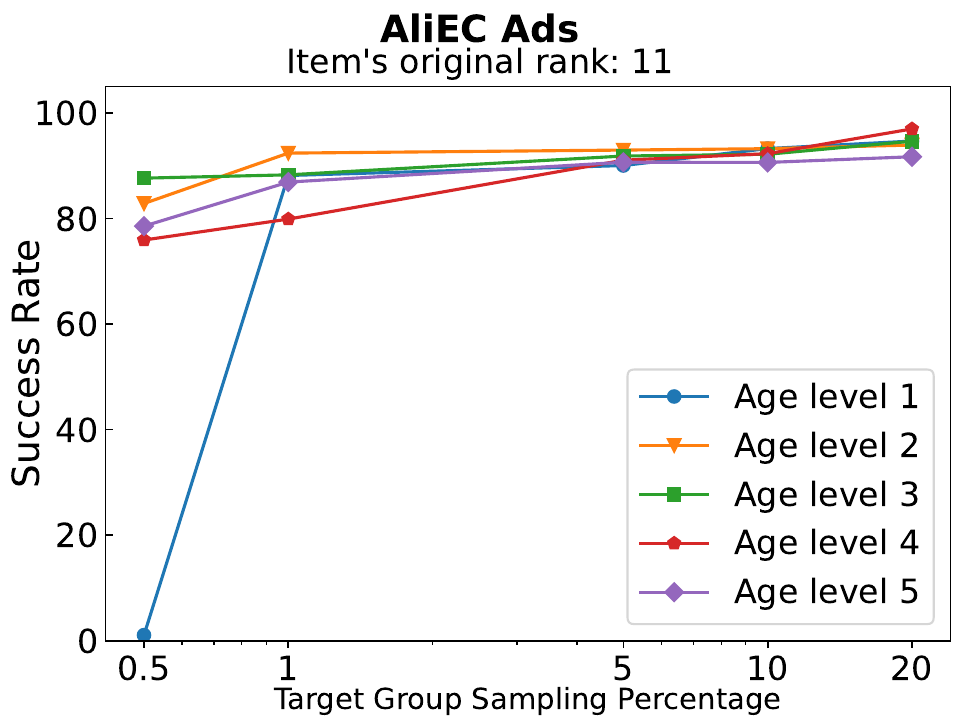}
    \includegraphics[width=0.5\columnwidth]{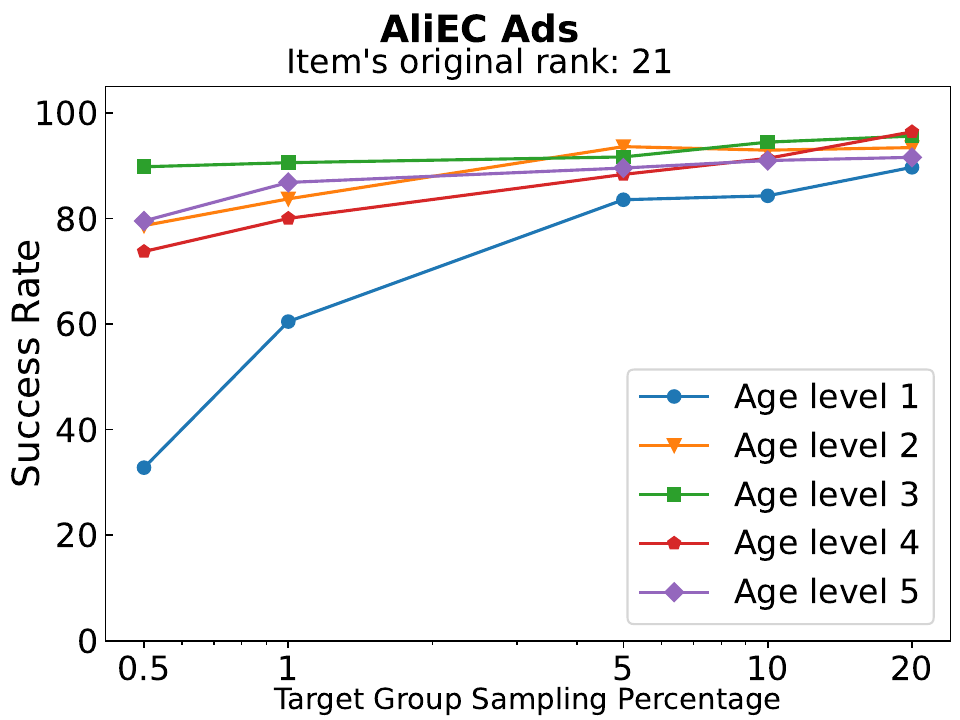}
    \includegraphics[width=0.5\columnwidth]{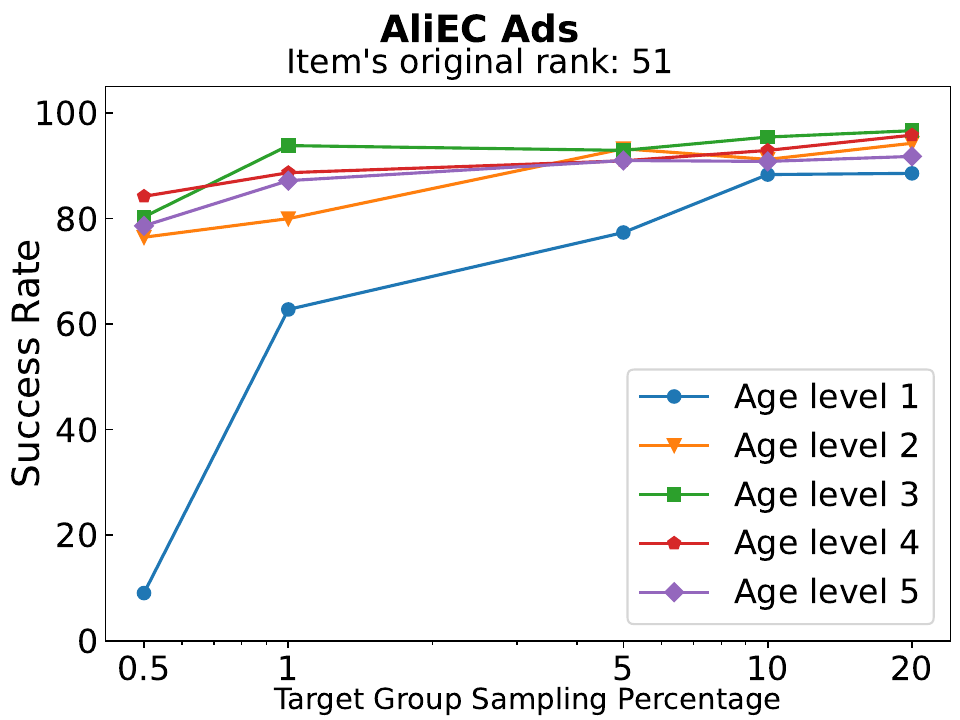}
    \includegraphics[width=0.5\columnwidth]{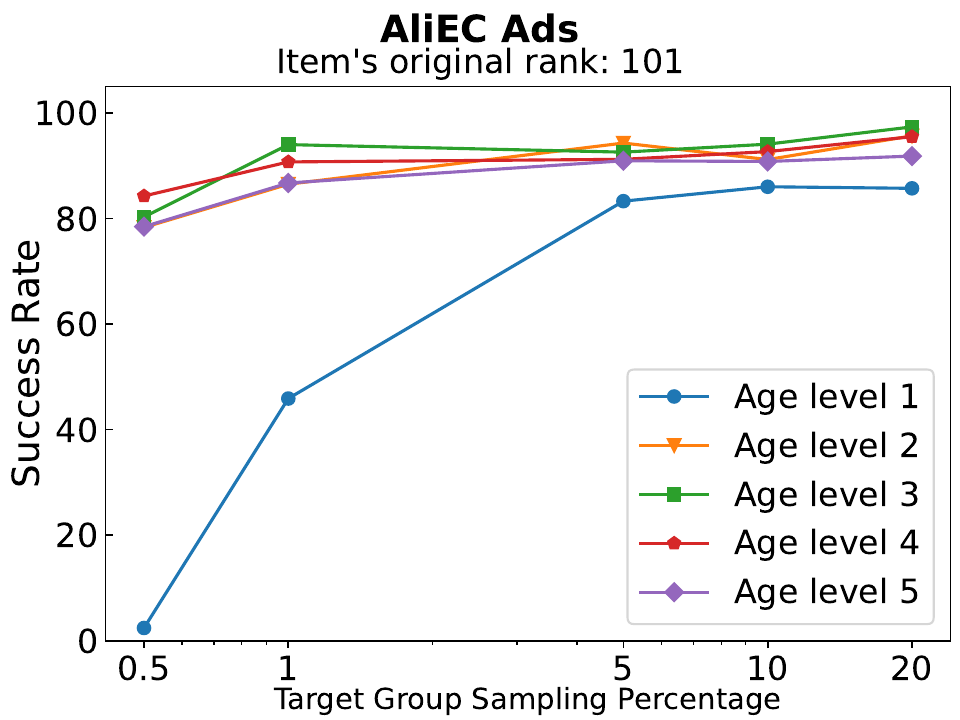}
    \caption{\small \toolname's success rate for items at original ranks 11, 21, 51, and 101 for the recommender system trained on AliEC Ads. \toolname gets more than 80\% success for most user groups and items at all original ranks with very small sampling size starting from 1\% of the user group, and increases to 100\% for with 20\% sampling. }
    \label{fig:aliec_ads_content_recsys}
\end{figure*}

\begin{figure*}
    \centering
    \includegraphics[width=0.5\columnwidth]{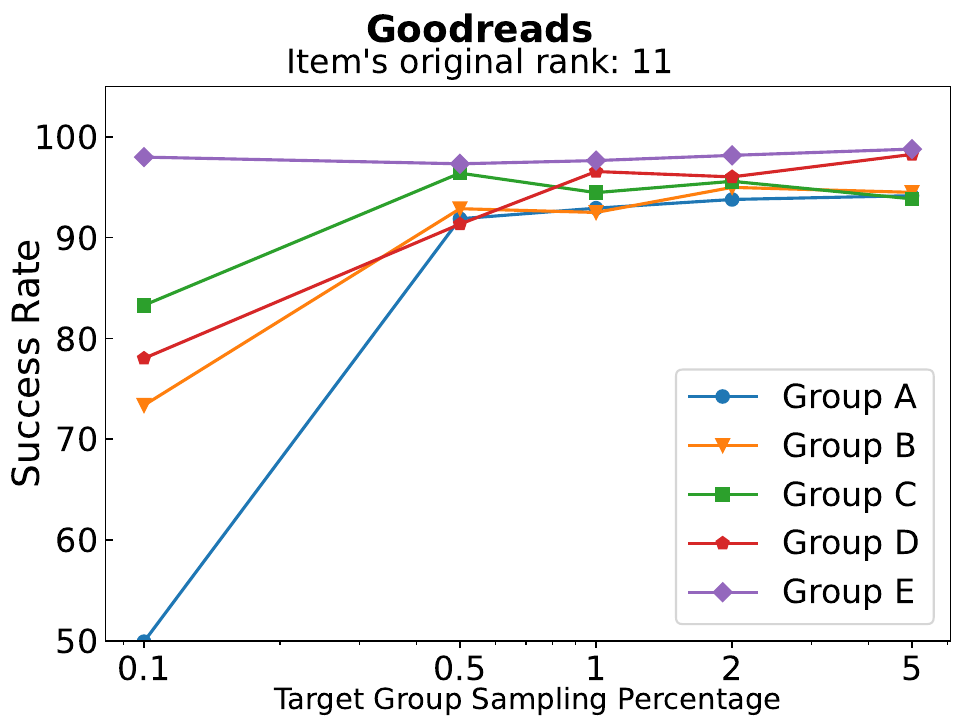}
    \includegraphics[width=0.5\columnwidth]{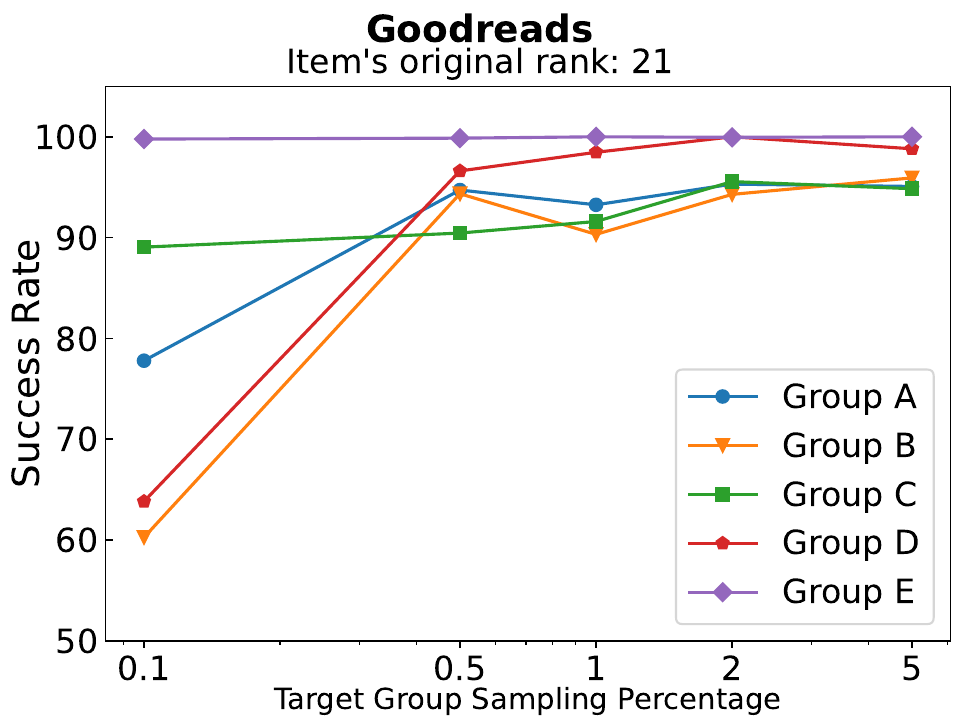}
    \includegraphics[width=0.5\columnwidth]{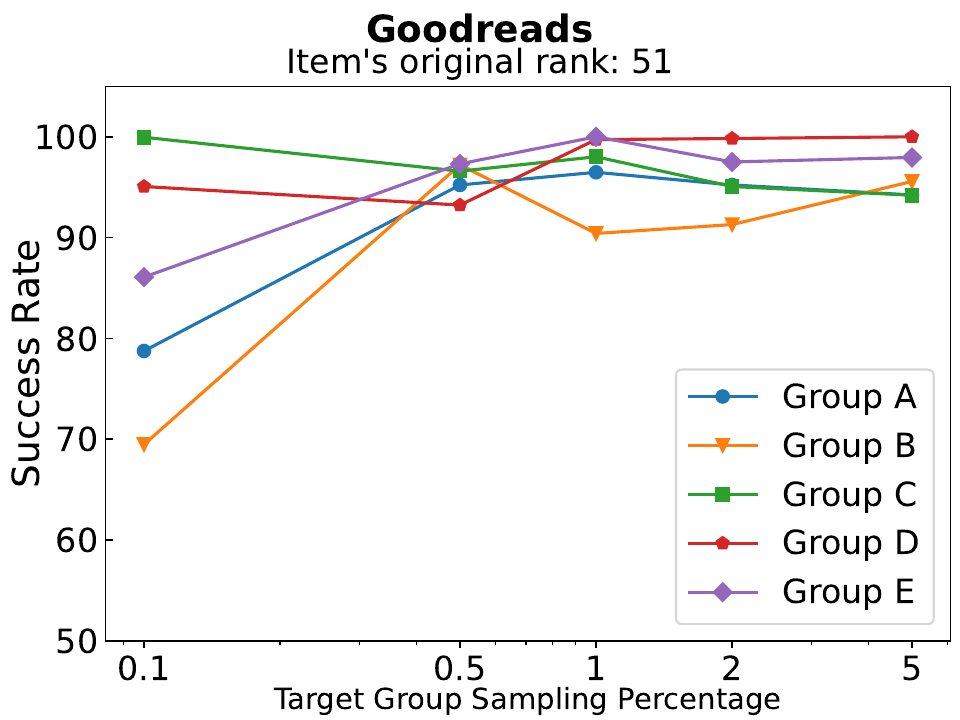}
    \includegraphics[width=0.5\columnwidth]{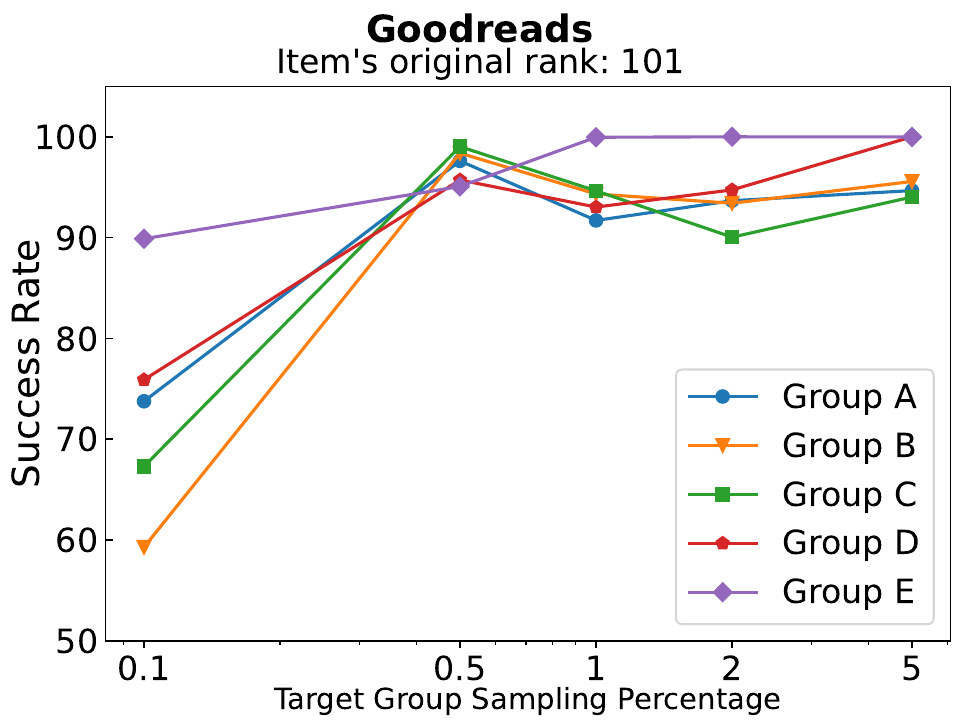}
    \caption{\small \toolname's success rate for items at original ranks 11, 21, 51, and 101 for the recommender system trained on Goodreads. \toolname gets more than 90\% success for all user groups and items at all original ranks with very small sampling size starting from 0.5\% of the user group, and increases to 100\% for with 2-5\% sampling. }
    \label{fig:goodreads_content_recsys}
\end{figure*}

\subsection{Experimental Methodology}
For all three recommender systems, we group the users into 5 equi-sized group either based on available metadata (like age) or based on the number of ratings they provided. Our experimental setup portrays a scenario where a content provider wants to increase the exposure of their item to a group of users, the \textit{target group}. 
\toolname provides a recourse to the content provider and after its execution, if the rank of their item improves to being within the top-$k$ recommendations for users in the target group, we would say that the item's exposure has increased and the provided recourse was valid. 
In experiments, we consider an item to have increased exposure for a user if after the recourse it is within the top-10 recommendation for that user. 
Even though the content providers would want to target a large group of users, optimizing the losses in \cref{eq:content_recourse_final} for all users in the target group can be computationally expensive. 
Therefore, we experiment using a sampling procedure. We randomly sample a small percentage of the users in the target group and minimize the loss only over them. 
However, we report the percentage of users in the \textit{entire target group} that get an increased exposure to the item. 
Intuitively, if we sample more users from the target group, a higher percentage of users would see the item within their top-10 recommendations. 

\begin{figure*}
    \centering
    \includegraphics[width=0.8\textwidth]{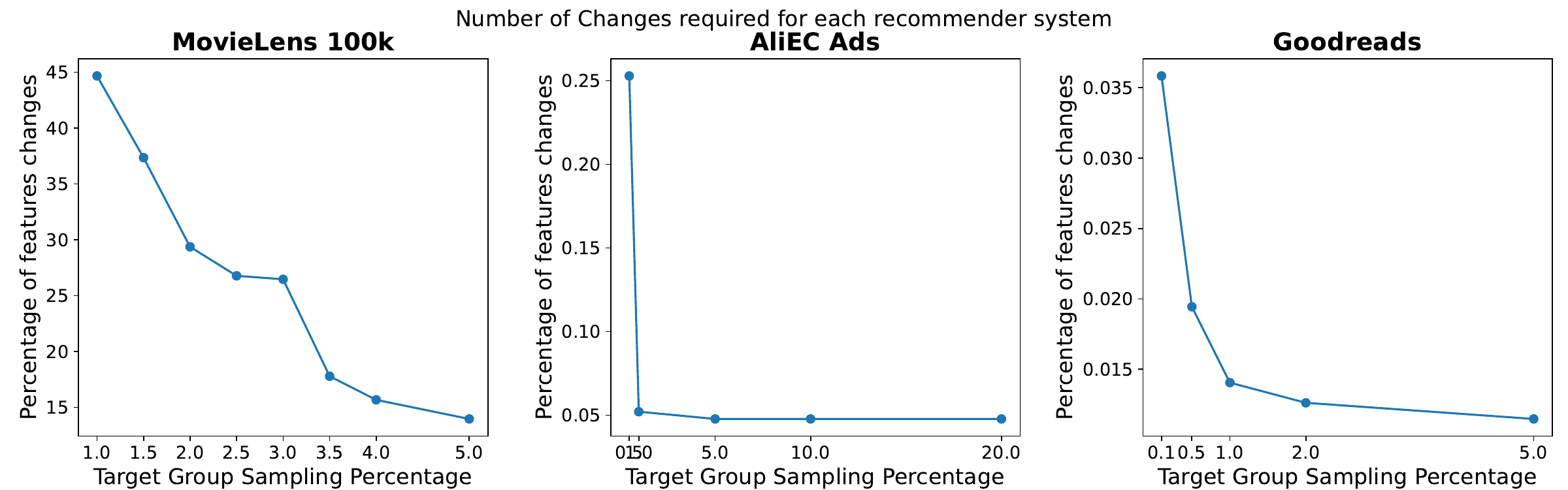}
    \caption{\small The percentage of item's features that need to be changed to execute the recourse. The plots are for the recommender systems trained on MovieLens-100K, AliEC Ads, and Goodreads. With increasing sampling percentage, the number of changes required to get a recourse decreases, and eventually becomes negligible. } 
    \label{fig:number_of_changes_required_content_recsys}
\end{figure*}

\textbf{Metrics. }
We report the following metrics for \toolname: 
\begin{enumerate}[leftmargin=9pt,topsep=0pt,partopsep=0pt]
    \item \emph{Success Rate: } For each recommender system, we report the percentage for users from the target group who see the concerned item withing their top-10 recommendations. We report this various sampling sizes. A higher value for this metric is better. 

    \item \emph{Number of changes required: } 
    This metric is computed as the L0 norm of the difference between the original features and the new features after recourse. A lower value for this metric is better. 
    
    \item \emph{Side-effect on user recommendations: } 
    To estimate the impact of a recourse on the users' original recommendations, we measure the similarity between the them and the new recommendations. We use rank-biased overlap (RBO) as a measure of similarity between the two ranked lists. 
    Since the change in recommendations at top ranks matter more, we weigh them more when measuring RBO similarity (p = 0.5). A higher value for this metric is better. 
\end{enumerate}

All the metrics are reported for items at various original ranks. Specifically, we generate recourses for items whose original ranks are 11, 21, 51, and 101 (item rank averaged over the users in the target group). For rigorosity, we ensure that an item whose rank is already within top-10 for more than 1\% of the users in the target group is not considered for getting a recourse. 

\Cref{sec:discussion} discusses the results from the experiments and \Cref{sec:conclusion} concludes the paper. 


\bibliographystyle{ACM-Reference-Format}
\balance
\bibliography{refs}

\clearpage
\appendix
\section{RESULTS AND DISCUSSION}
\label{sec:discussion}




\Cref*{fig:ml_100k_content_recsys,fig:aliec_ads_content_recsys,fig:goodreads_content_recsys} show the success rate of \toolname for items at different original ranks for the recommender systems trained on MovieLens-100K, AliEC Ads, and Goodreads respectively. We see that for all the three recommender systems, \toolname achieves very high success rate (more than 80\% for all datasets) at very low sampling of 1\% or less of the target group. The success rate improves to 100\% for all datasets when the sampling percentage is increased to 5-20\% depending on the dataset. 

\Cref{fig:number_of_changes_required_content_recsys} shows the percentage of item features that need to be changed to execute the recourse. Across all the three recommender systems, user groups, and original item ranks, the number of changes required to get a recourse decreases with increasing sampling percentage, and eventually becomes negligible. 

The similarity between the original and new recommendations for all users was consistently very high. For MovieLens-100K, the average RBO metric (across user groups, original ranks, and sampling percentages) was 99.43\%, for AliEC Ads, it was 99.99\%, and for Goodreads, it was 99.92\%. Therefore, the recourses generated by \toolname do not change the user's recommendations significantly. 

Therefore, the metrics demonstrate that \toolname is able to generate valid recourses for recommender systems with a small sampling of the target group. The recourses it generates do not require a large number of changes to the item's features, and the new recommendations for the target users do not change significantly from the original recommendations. 

\section{CONCLUSIONS}
\label{sec:conclusion}

We propose a novel approach called \toolname that generates recourses targeted towards content providers in a content filtering-based recommender system platforms. Previous research has established how important it is for the content providers to understand the factors that influence the ranking of their product on the recommender system based platforms \citep{airbnbhosts-xaineed,Upwork-xaineed,FB-seller-xaineed,Etsy-xaineed}, and thereby improve their product's ranking. 
To the best of our knowledge, \toolname is the first approach to generate recourses for recommender systems. 
Experimental results with three datasets used to train the recommender systems demonstrate \toolname's ability to generate recourses that satisfy the desiderata (\Cref{sec:desiderata}) of a recourse. 
It generates recourses that have a high success rate, require a small number of changes to the item's features, and cause little to no side-effect on the target users' recommendations. 

\begin{appendices}
\section{RELATED WORK}
\label{sec:related}

Machine Learning (ML) is being increasingly used to automate decisions. Some of the applications where ML is being used are highly critical 
and directly affect humans, for example, loan approval \citep{credit-risk-ml}, criminal justice \citep{parole-ml}, and hiring \citep{hiring-ml}. 
The nascent field of trustworthy ML aims to detect bias in ML models (and counteract it), understand the factors that the ML model is using in making predictions, 
ensure the models respect privacy and security, and frame policies and regulations that the ML models should abide by \citep{barocas-hardt-narayanan,trustml1,trustml2}. 
Research has established a few dimensions of trustworthy ML, like, fairness \citep{fairness1,fairness2,barocas-hardt-narayanan,fairness4,fairness5,fairness6,fairness7}, 
interpretability \citep{interpretable1,interpretable2,interpretable3,interpretable4,interpretable5,interpretable6,interpretable7,interpretable8}, 
robustness \citep{robustness1,robustness2,robustness3,robustness4,robustness5,robustness6,robustness7,robustness8}. 
In this work, we focus on interpretability and refer the readers to work by \citet{barocas-hardt-narayanan} and \citet{varshney} for a broad discussion of trustworthy machine learning. 

\subsection{Interpretability in ML}
Interpretability is the branch of trustworthy ML that aims to provide human consumable explanations for the predictions made by ML models 
that are used for tasks such as classification, regression, and recommendation. Most of the research in this area has focused on the interpretability of classification models. 
Interpretability for classification models can be achieved by either developing inherently interpretable ML models 
(e.g., logistic regression, shallow decision trees) or a post-hoc explanation of complex ML models (e.g., random forests, neural networks). 
Post-hoc explanations can be further bifurcated into generating feature attributions using techniques like SHAP \citep{shap} or generating counterfactual explanation-based recourses \citep{wachter2018counterfactual}. 
Feature attribution explanations highlight the features that might have been important in making a prediction. 
On the other hand, counterfactual explanations provide a counterfactual situation that would have led to a different prediction from the ML model. 
\citet{Miller-xai:2019} in a social science study remarked that when people ask `Why P?' questions, they are typically asking `Why P rather than Q?', 
where Q is implicit in the context of the application. An example of this case is the question a person whose loan request has been rejected would ask: 
`Why has my loan request been rejected?', which actually means: `Why has my loan request been rejected instead of being accepted?'. 
And counterfactual explanations are a way to answer this question. They would respond, for example, by saying that `had your income been \$3000 higher, you would have gotten the loan'. 
This simultaneously also provides a recourse to the affected individual, who now knows that they can get the loan if they can increase their income by \$3000.

\subsection{Interpretability in Recommender Systems}
Literature in interpretability for recommender systems has focused on highlighting the factors that might have contributed to a recommendation. 
This is similar to feature attribution based explanations for classification models. 
Interpretability research for recommender systems can be categorized into user-based, item-based, and feature-based explanations. 
In user-based explanations, a high rating for the item provided by a group of users similar to the user is given as an explanation for the recommended item. 
In item-based explanations, the recommended item is explained by its similarity to the items that the user has liked or purchased in the past. 
Feature-based explanations highlight the features of the recommended item that the user has shown interest in the past, for example, the cast for movie recommendations. 
The approaches that generate these explanations can either be model-specific or model agnostic. 
We refer the readers to a survey on the explainability of recommender systems for a more comprehensive discussion \citep{survey-reco-xai} on this topic. 
Similar to feature attribution explanation for classification models, a noticeable characteristic of the aforementioned explanations for recommender systems is that they are not actionable. 

\toolname, on the other hand, generates recourses targeted towards the content providers of the recommender systems. This is the main contribution of this work. 
It provides counterfactual features that would lead to a different ranking of a specific item in the recommendation list for a target group of users.  

\subsubsection{Previous studies on need for recourse for content providers}
\label{sec:need-for-recourse}
In this subsection we continue to discuss related work that has highlighted the need for recourse for content providers. 
\citet{airbnbhosts-xaineed} did a study with several Airbnb hosts to understand their perspectives. 
They clearly expressed the need for transparency and recourse on the platform. 
One of the hosts said: \textit{``I feel less motivated because I don't think that it's clear what I need to do, and I think that it's frustrating seeing the search: 
lots of listings that are worse than mine are in higher positions.''}
Several hosts performed A/B testing with different factors like pricing adjustments, calendar updates, location, type of room, 
amenities like free parking, changing descriptions of the property, allowing dogs, allowing short-term vs. long-term guests, etc., to understand 
which factors can help improve their ranking. 
\citet{Upwork-xaineed} interviewed freelancers working on Upwork. Freelancers also struggled in understanding what factors go into the ranking and how they can influence it. 
One of the freelancers said: \textit{``all I can think about is figuring out how to raise my score''}. 
Similar to Airbnb hosts, freelancers tried and tested changing different attributes of their profile, like taking technical tests provided by Upwork, 
opening and closing contracts, having shorter projects, and inflating the hourly working rate to improve their ranking. 
A precisely similar need and behavior was observed when studies were conducted with sellers on Facebook MarketPlace \citep{FB-seller-xaineed}, 
freelancers on other platforms like TaskRabbit and Fiverr \citep{Upwork-course1,Upwork-course2}, drivers using Uber and Lyft \citep{Uber-xaineed}, 
and sellers on handmade product platform Etsy \citep{Etsy-xaineed}. 
Several freelancers who \citet{Upwork-xaineed} interviewed mentioned that owing to the black-box behavior of the algorithmic freelancing platform, 
they made frequent efforts to take the work offline or pause the work to pacify the algorithm, and some even quit. 
\citet{Upwork-course1} had the same observations in their interviews with Upwork's freelancers. 

Providing recourses to content providers of a platform would help them understand what actions they can take to improve their product's ranking and get transparency into the current ranking. 


\subsubsection{Counterfactual Explanations in Recommender Systems}
There have been recent proposals for some approaches that seek to generate explanations for recommended item in a counterfactual manner 
\citep{cfe-reco-approach1,cfe-reco-approach2,cfe-reco-approach3,cfe-reco-approach4,cfe-reco-approach5}. 
All these approaches explain a recommendation by finding the smallest change in the user's interaction history that would replace the top recommendation with anything else. 
For example, an explanation for the top-recommended movie \emph{The Godfather II} is that the user had previously liked \emph{Goodfellas} and \emph{The Godfather}. 
Had the user not liked these two movies, \emph{The GodFather II} would not be the top recommendation (it could still be recommended but at a different rank). 
\toolname is distinct from these works in several ways:
\begin{itemize}[leftmargin=*,topsep=0pt,partopsep=0pt]
 \item \toolname provides recourse to the content providers of the recommender system, while the aforementioned approaches provide explanations to the users of the recommender system. 

 \item \toolname provides a set of actions that can be executed to get a favorable rank for an item while the aforementioned 
 approaches do not provide that. They only find the smallest change that would replace the top recommendation with anything else, not something the content provider or the user wants. 
 
 \item \toolname can make suggestions to change features that are \textit{not in} the user's history, while the aforementioned approaches only alter the user's history. 
 
 \item \toolname is able to provide a recourse for items at any rank (in order to get them to an improved rank) in the recommended list, while the aforementioned approaches provide an explanation for only the top-ranked item. 
\end{itemize}

Our work also has subtle similarity to the work by \citet{sarah_rec}, where they define reachability as the feasibility of the end-user of a recommender system
modifying their rating in order to get an item recommended. 
\toolname is distinct from \citet{sarah_rec}'s work in the following ways:
\begin{itemize}[leftmargin=*,topsep=0pt,partopsep=0pt]
    \item Their work is concerned with only the end-user of a recommender system, while \toolname is targeted towards the content providers of the recommender system. 

    \item The goal of their work is to audit the recommender system to understand whether it could cause polarization or filter bubbles, 
    while the goal of \toolname is to provide recourse to the content providers of the recommender system. 

    \item Their approach is limited to matrix factorization based recommender systems, while \toolname generalizes not only to all architectures of collaborative filtering recommender systems, but also to content based recommender systems. 
\end{itemize}
\end{appendices}

\end{document}